%% file: preprint.tex
\documentclass[
hf,
]{ceurart}

\usepackage{listings}
\usepackage{csquotes}
\usepackage{enumerate}
\usepackage{paralist}
\usepackage{orcidlink}
\usepackage{rorlink}
\usepackage[defaultlines=5,all]{nowidow}

\begin{document}

\copyrightyear{2024}
\copyrightclause{Copyright for this paper by its authors. Use permitted under Creative Commons License Attribution 4.0 International (CC BY 4.0).}

\conference{This paper has been submitted to PLOS ONE in January 2024 and was desk rejected as it is more a policy/method paper. It has been resubmitted to F1000Research}

\title{Towards a Quality Indicator for Research Data publications and Research Software publications \protect\\– A vision from the Helmholtz Association}
\shorttitle{Towards a Quality Indicator for Research Data and Software Publications}

\author[1]{Wolfgang {zu Castell}}[orcid=0000-0003-2820-6564]
\author[1]{Doris Dransch}[orcid=0000-0003-0839-8167]
\author[2]{Guido Juckeland}[orcid=0000-0002-9935-4428,email=g.juckeland@hzdr.de]\cormark[1]
\author[3]{Marcel Meistring}[orcid=0000-0001-6347-9926]
\author[4]{Bernadette Fritzsch}[orcid=0000-0002-0690-7151]
\author[5]{Ronny Gey}[orcid=0000-0003-1028-1670]
\author[6]{Britta Höpfner}[orcid=0000-0001-6159-4875]
\author[7]{Martin Köhler}[orcid=0000-0003-0617-3319]
\author[1]{Christian Meeßen}[orcid=0000-0001-8151-8722]
\author[8]{Hela Mehrtens}[orcid=0000-0002-4526-2472]
\author[1]{Felix Mühlbauer}[orcid=0000-0002-0727-8326]
\author[9]{Sirko Schindler}[orcid=0000-0002-0964-4457]
\author[5]{Thomas Schnicke}[orcid=0000-0001-9868-4671]
\author[3]{Roland Bertelmann}[orcid=0000-0002-5588-0290]

\address[1]{Helmholtz Center Potsdam - German Research Center for Geosciences, Potsdam, Germany~\rorlink{https://ror.org/04z8jg394}}
\address[2]{Helmholtz-Zentrum Dresden-Rossendorf (HZDR), Dresden, Germany~\rorlink{https://ror.org/01zy2cs03}}
\address[3]{Helmholtz Association, Helmholtz Open Science Office, Berlin, Germany~\rorlink{https://ror.org/0281dp749}}
\address[4]{Alfred Wegener Institute, Helmholtz Center for Polar and Marine Research, Bremerhaven, Germany~\rorlink{https://ror.org/032e6b942}}
\address[5]{Helmholtz Centre for Environmental Research, Leipzig, Germany~\rorlink{https://ror.org/000h6jb29}}
\address[6]{Helmholtz-Zentrum Berlin für Materialien und Energie (HZB), Berlin, Germany~\rorlink{https://ror.org/02aj13c28}}
\address[7]{Deutsches Elektronen-Synchrotron DESY, Hamburg, Germany~\rorlink{https://ror.org/01js2sh04}}
\address[8]{GEOMAR Helmholtz Centre for Ocean Research, Kiel, Germany~\rorlink{https://ror.org/02h2x0161}}
\address[9]{Institute of Data Science, German Aerospace Center (DLR), Jena, Germany~\rorlink{https://ror.org/04bwf3e34}}

\cortext[1]{Corresponding author.}

\input{tex/0_abstract}

\maketitle

\input{tex/1_content}

\begin{acknowledgments}
  \input{tex/2_acknowledgment}
\end{acknowledgments}

\appendix
\input{tex/3_credit}

\pagebreak
\bibliography{bibliography}

\end{document}

%% file: tex/0_abstract.tex
\begin{abstract}
Research data and software are widely accepted as an outcome of scientific work. 
However, in comparison to text-based publications, there is not yet an established process to assess and evaluate quality of research data and research software publications. 
This paper presents an attempt to fill this gap. 
Initiated by the Working Group Open Science of the Helmholtz Association the Task Group Helmholtz Quality Indicators for Data and Software Publications currently develops a quality indicator for research data and research software publications to be used within the Association. 
This report summarizes the vision of the group of what all contributes to such an indicator. 
The proposed approach relies on generic well-established concepts for quality criteria, such as the FAIR Principles and the COBIT Maturity Model. 
It does -- on purpose -- not limit itself to technical implementation possibilities to avoid using an existing metric for a new purpose. 
The intention of this paper is to share the current state for further discussion with all stakeholders, particularly with other groups also working on similar metrics but also with entities that use the metrics. 
\end{abstract}

%% file: tex/1_content.tex
\section{Background and objective}

Measuring the performance and progress of research based on means of output is a common task of research assessment. 
While it is commonly accepted to use quality-assured scientific publications as a proxy for quality assessment in research evaluation, other artifacts of scientific work such as research data and research software are often not yet considered.
For text-based scientific work, there are well established processes of quality assurance during the process of publication. For publishing research data and research software no commonly accepted standards exist yet, although research data and software represent a significant output of the scientific endeavor especially given the rise of data- and software-intensive research.

The Helmholtz Association has already decided to integrate research data publications and research software publications in its internal key performance indicator system and is currently developing and implementing a quality indicator for those artifacts. 
It should become a driver within Helmholtz in a threefold way: 
a) to create awareness and appreciation for the diversity of research output, 
b) to align the assessment of research and research practice to the conditions of digitalization and openness of science, and 
c) as a stimulus and incentive for scientists to improve their data and software publications related to specific quality criteria, and thus, to improve the quality and trustworthiness of science as a whole. 
Helmholtz Centers from all research fields of the Association contributed in an iterative discussion process to cover a broad perspective originating from their respective research environment. 
To meet all aspects relevant for the quality indicator, a group of scientists, scientific software developers, data curators, and colleagues reporting on research output were involved in its development. 
This paper provides insight to the current status of discussion.

\section{Terminology}

The following definitions should clarify what type of research data and research software we address in this paper.

\begin{description}
    \item[Research data]
        In accordance with the EOSC-Glossary~\cite{EOSCGlossary} we define research data as data that has been collected, created and/or is being used in \enquote{scientific research activities and used as evidence in the research process, or commonly accepted in the research community as necessary to validate research findings and results}.
        It ranges from highly complex and individual long-tail-data from experiments and observations to automatically produced generic data collections.
    \item[Research software]
        We define research software as software that has been developed in the research process~\cite{gruenpeter2021}.
        We consider software of all levels of maturity:  from personal use, reuse by others, long term support, and criticality as defined by \cite{Schlauch2018} in the form of software classes.
    \item[Published data and software]
        In both cases we concentrate on research output which is published. 
        We define Published data and software as digital objects that have been made as permanently available as possible on the Internet~\cite{Lawrence2011}.
        The different levels of “Published” are addressed in the framework of quality dimensions and attributes (\autoref{sect:data} and \autoref{sect:software}).
\end{description}

\section{Methodical framework – quality criteria and quality quantification}

Recalling the publishing process of scientific publications: 
A paper published e.g. in a scientific journal has been quality checked with respect to several criteria over the course of the publication process. When accepted, a manuscript is marked as being approved by the underlying quality assurance mechanisms. 
Aggregating the number of papers that went through this process of quality assurance is therefore used as a quantitative measure of the output performance of a scientific entity. 
Yet, this process of solely using simple and one-dimensional metrics to assess research quality and performance is broadly criticized. 
But the convenience of having one number easy at hand has so far hindered all attempts to replace this practice. 
Taking this into consideration a quality indicator for research data and software has to be developed with the goal to have one easily comparable value that still does more then only counting - by representing a diverse set of quality features in itself.

Therefore, any analog of this process for scientific data publications or research software publications has to technically map a multi-dimensional suite of quality criteria onto one value, the quality indicator.
Thus, developing a quality indicator has to address the following questions: 
1) What quality criteria should be considered?  
2) How to measure each of the quality criteria? 
3) How to condense individual quality measures into one number, the quality indicator? 
In addition, for practicability, 4) how to collect the quality measures for the most part in an automated fashion? 
Since the Helmholtz quality indicator has to match all types of research data publications and research software publications generated in the diverse scientific disciplines within the Helmholtz Association, generic approaches are required to provide answers to these questions. 
We decided to build on the \textit{FAIR Principles}~\cite{Wilkinson2016} to specify relevant quality criteria. 
In order to provide a generic way to estimate quality for a wide range of criteria, we are using an approach based on the \textit{COBIT Maturity Model}~\cite{DeHaes2015} to quantify quality in terms of a maturity scale. 
Finally, a \textit{weighted summation and radar plot} approach is used to derive one single value from the various quantitative measures while also providing a more detailed representation. 
The collection of quality measures should rely on \textit{automated procedures and tools} as much as possible. 
However, in the end, any indicator must match the target set, technical issues must not lead to deviation from that target.
Therefore, our basic approach for developing the quality indicator was to define the quality criteria that are essential for the intended objectives in a first step, and to then condense the list of criteria in a second step with respect to technical feasibility of automated collection. 
This approach discusses quality criteria of research data publications and software publications without immediately considering the technical implementation of measuring them. 
Thus, we selected metrics not simply for the ease of measurement. 
Rather, the measurement of some criteria deemed highly relevant might be deferred until a corresponding, automated approach of evaluating them becomes available.

The overall approach then consists of the following five steps.

\begin{enumerate}[(i)]
    \item Define a suitable number of dimensions, according to which quality of research data and software can be assessed.
    \item For each of these dimensions, collect a set of specific attributes capturing the respective dimension.
    \item Using a generic maturity model, each attribute is mapped onto a numeric scale 
    \item Using weighted averages, an aggregated maturity level is obtained for each dimension.
    \item The summarizing quality assessment can then be visualized using radar plots.
\end{enumerate}

These steps lead to the current state of a quality summary of the assessed research data and software.
Using more than one dimension allows us to take different perspectives into account. 
For example, an organization aiming at building-up a large, integrated software library might have other targets for quality assessment than a single software developer working on individual projects. 
As well as an international campaign gathering huge amounts of automatically generated sensor-data has other objectives, restrictions, and conditions than long-tail data generated in experiments in a laboratory setting or a medical scientist working with confidential patient data. 
In analogy to the aggregation of quality-assured scientific publications, a quantitative estimate of high quality data and software publications could then be obtained by aggregating the number of data/software showing a maturity above a given minimum in each of the dimensions.
Afterwards, the maturity of each dimension also has to be aggregated to form the final indicator for a publication.

To determine an initial set of dimensions for quality assessment, we are building on the FAIR principles~\cite{Wilkinson2016}.
The FAIR principles have been widely adopted as a minimum set of quality criteria for research data and research software  thus, our approach generally also builds on the four principles findability, accessibility, interoperability and re-usability. 
Considering upcoming discussions within communities on the limitations of pure implementation of FAIR to grasp a comprehensive view on the quality of research data and research software publications, the partly interrelated aspects and minor inaccuracies of wording within the four core principles we adopted and extended the FAIR principles to best suit our needs.

For research software we amended the FAIR criteria with two additional criteria: \textit{scientific quality}, addressing the question of research software being scientifically well grounded, and \textit{technical competence}, capturing aspects of the software being technically well grounded (FAIR-ST).

For evaluating research data publications, we regrouped the FAIR criteria into four dimensions that are easier to grasp related to datasets but still cover the same aspects. 
Those being 1. \textit{Publishing}, 2. \textit{Openness}, 3. \textit{Curation}, and 4. \textit{Metadata}. 
To compensate for inadequacies between different disciplines when applying a general framework like the proposed, the 5. dimension of the \textit{External View} has been added to the set (POCME). 

In order to allow for a comparable and unbiased quality assessment for data and software publications based on our modified FAIR-framework, we need a broadly applicable scheme leading to a numerical value for each quality criterion.
In certification processes, maturity models are commonly used to solve this task.

\subsection{The COBIT Maturity Model}

To achieve comparability among different attributes of quality, stages of product evolution can be assessed using a generic model of maturity.
Being generic, the model can be applied to all categories of scientific output leading to quantitative results which can be compared.
Our approach utilizes the COBIT Maturity Model~\cite{DeHaes2015}.
The COBIT Maturity Model quantifies maturity of a product in six levels by mainly evaluating the processes that lead to the product.
Looking at processes for achieving a certain quality is a common approach in quality management, where quality management systems (QMS) are defined as \enquote{formalized systems that document processes, procedures, and responsibilities for achieving quality policies and objectives}~\cite{ASQ}.
Typically, lower levels of maturity focus on individual outcomes of a process to be certified.
The higher the maturity, the more the process is considered, accounting for the fact that a well-developed process naturally leads to products of a certain quality.
COBIT starts with a default level 0 (non-existent), two levels of maturity acknowledge the successive completion of a set of quality criteria for the product to be targeted.
Starting with level 3 maturity implies a process to be existent which guarantees the addressed quality to be kept.
Two further levels finally focus on the improvement of the underlying process towards full optimization with maturity level 5.
Clearly, the explicit meaning of the maturity levels for a given quality criterion has to be defined as part of the community process.
Obviously, the maturation levels need to be interpreted within the context of a given aspect of quality assessment.
The levels are cumulative in the sense that a higher level includes the requirements of all lower levels, i.e. marking a higher level as fulfilled means that the requirements of all lower levels are fulfilled plus the requirements of the highest marked level.
Thus, a higher level can either refine/specialize characteristics that were already evaluated for compliance with a lower level or introduce a new and additional characteristic to assess. As a result, comparability across quality dimensions is given.

The specific application of maturity levels derived from the COBIT Maturity Model with regards to data and software publications are introduced in the respective subsection of this paper (\autoref{sect:data:indicators} and \autoref{sect:software:indicators}).

\subsection{Weighted summation and radar plot}

Guidelines for reviewers for academic journals typically provide a set of criteria being considered in preparing a review for a manuscript.
Examples are scientific originality, proper use of methods, organization of the material/figures/tables, and comprehensibility.
Such criteria can likewise be grouped into different dimensions of assessment such as scientific novelty, being scientifically grounded, quality of presentation etc.
However, such dimensions are commonly not transparently weighted against each other. Nevertheless, there is a common understanding in each scientific community on what has to be considered more important.
To make such an understanding open and transparent, we introduce radar plots to visualize the joint maturity of a data and software publication with respect to all quality dimensions.
The relative importance of every single attribute within each dimension of assessment can be fine-tuned by agreeing on weights to be used in a weighted average.
Doing so, minor criteria can also be included into the assessment without the risk of leading to a biased assessment. 

\begin{figure}[!hb]
    \centering
    \includegraphics[width=0.7\textwidth]{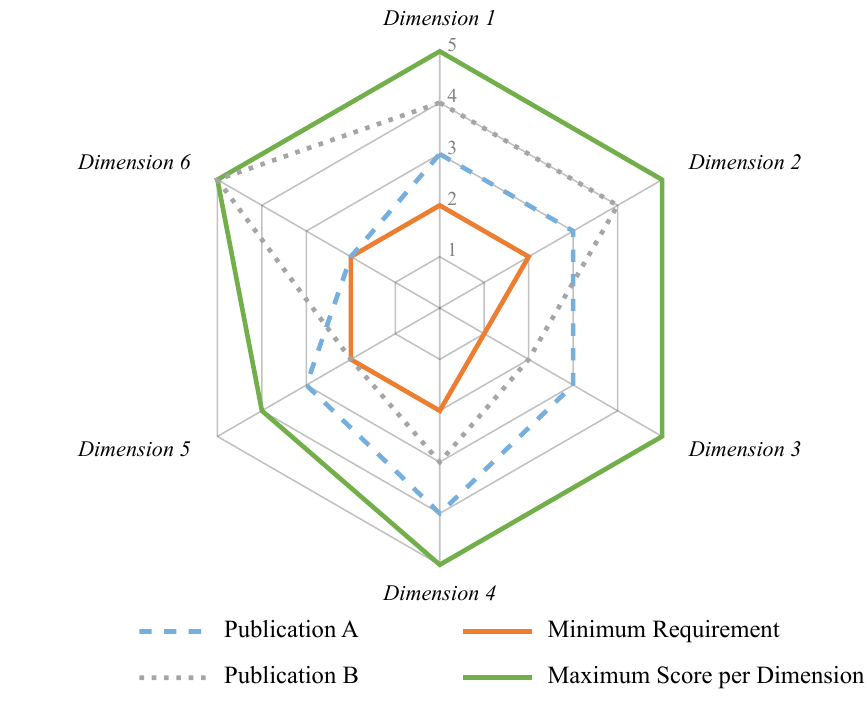}
    \caption{
        Example of a fictitious assessment for two research  software or data publications showing how the radar plot can visualize the fulfillment of some criteria at higher quality levels than others.
        Maturity levels are ordered from center (0, non-existent) to outside (5, optimized). The minimum requirements and maximum scores might also differ for each dimension.
    }
    \label{fig:radarPlot}
\end{figure}

For the overall assessment of the data and software publication, we deliberately do not suggest to use weighted averages as a default.
The multi-dimensionality of each quality assessment should rather be kept transparent.
This allows to optimize towards different goals in using the quality indicator.
\autoref{fig:radarPlot} exemplifies the result of the quality assessment of two fictitious research software or data publications.

\section{The Quality Indicator for research data}
\label{sect:data}

\subsection{Dimensions of quality for research data assessment}

The quality of published research data and hence its potential value for future reuse in other contexts can be assessed based on the quality of the metadata describing the data formally, on metadata describing the data on the content-level, on the quality of the data itself and the conditions of Openness/Licensing. 
With regards to the diversity of research data over the different scientific domains represented at Helmholtz, the heterogeneous ways of collecting, creating, editing, and curating data, a single quality criterion for research data assessment would not be constructive.
To allow for a multi-dimensional perspective on research data, the quality assessment being presented in this document is spanning five dimensions of data quality being partly based on the FAIR-principles and the FAIR Data Maturity Model of the Research Data Alliance~\cite{RDAFDMMWG2020}.

The derived quality dimensions of the POCME-framework for research data publications, each with indication of the covered FAIR-principles, are:

\begin{description}
    \item[Publishing]
        This dimension covers especially the areas of Findability and Accessibility. 
        It addresses the use of identifiers to enable unambiguous identification as well as the level of storage and hence the possibilities and information on the general accessibility to published data.
    \item[Openness]
        As publishing a dataset does not necessarily mean to make it openly available (Intelligent Openness), this dimension takes this circumstance into account and covers aspects of Accessibility, Interoperability and Reusability.
    \item[Curation]
        Curation can be crucial for the quality of published data itself as well as for the provided formal and content related information. 
        Hence, it can significantly improve the value of a published dataset for Reusability as well as for Findability, Accessibility, and Interoperability.
    \item[Metadata]
        Qualitative metadata can help in many aspects of Findability, Accessibility, Interoperability, and Reusability of published data. 
        By providing content related aspects, helpful information to retrieve a dataset as well as context around a dataset.
    \item[External View]
        This dimension is meant as a handle to compensate inadequacies between different disciplines which is a sensible approach when applying a general framework like the proposed.
\end{description}

These five dimensions of quality for research data are still too generic to be evaluated using the maturity model.
Therefore, each dimension is split into a set of attributes.
The latter will be assessed using the maturity model.
To derive a single value for each of the five dimensions, an aggregation step has to be incorporated.
This is done by computing a weighted average of the maturation values for all attributes constituting a quality dimension.
Using weights provides the freedom not to be forced to work with attributes of equal importance.
Minor aspects are just given smaller weight.

\subsection{Quality assessment of the dimensions}
\label{sect:data:indicators}

To assess the quality of research data publications, the developed POCME assessment-framework measures/estimates the maturity of different quality dimensions and attributes.
To be able to do this, a generic five-level scale of maturity derived from the COBIT Maturity Model is applied.
The levels are always to be seen in the context of the respective dimension and attribute they are applied on. 

The first level (0) usually implies that the respective attribute is non-existent or not applied.
The following levels built upon each other, always following the assumption, that the previous levels are largely met, leading to the highest possible level of maturity and hence quality of the data itself and the accompanying meta-information maximizing the findability, accessibility, interoperability, and re-usability.

\begin{enumerate}[(1)]
    \setcounter{enumi}{-1}
    \item Non-existent: no information available or not applied
    \item Most necessary information provided or measure taken
    \item Basic information provided or measure taken (sensible level of information/measures)
    \item Advanced information provided or measure taken, allowing to generally understand and (re)use the published data
    \item Complete and accurate information provided or measure taken, to an extend that allows maximal understanding and usage of data
\end{enumerate}

The following sections list the attributes and their maturity levels of research data publications that are proposed for each dimension.
It must again be noted that the COBIT model is cumulative, i.e. a higher level includes the fulfillment of all requirements of a lower level.

\subsubsection{Dimension Publishing}

\paragraph{Published with Identifier}
\begin{compactenum}[(1)]
    \setcounter{enumi}{-1}
    \item No identifier (resource may only be found via personal communication)
    \item Basic Uniform Resource Identifier
    \item Dataset is identifiable via internal handle (does not resolve globally, generally no  metadata)
    \item Dataset is basically identifiable via formalized, standardized, persistent identifier (resolves globally, general metadata provided)
    \item Dataset is identifiable via globally unique, formalized, standardized, persistent  identifier supported by general metadata (e.g. DOI).
\end{compactenum}

\pagebreak
\paragraph{Published via a Repository or Collection, that is indexed in a Meta-Repository (e.g. re3data)}
\begin{compactenum}[(1)]
    \setcounter{enumi}{-1}
    \item No information available, the data is not published via a repository/collection
    \item The data is published in a repository/ collection which is not listed in an eligible meta-repository
    \item The repository/collection is listed in an eligible meta-repository, basic no. of quality indicators assigned by the meta-repository are achieved
    \item The repository/collection is listed in an eligible meta-repository, medium no. of quality indicators assigned by the meta-repository are achieved 
    \item The repository/collection is listed in an eligible meta-repository, high no. of quality indicators assigned by the meta-repository are achieved
\end{compactenum}

\bigskip\noindent
Remark: 
Meta-repositories like re3data usually do not perform in depth-checks on the quality of published research data in the indexed sources. Still, if sources indexed in meta-repositories like re3data fulfill certain quality-criteria respective indicators (e.g. icons) are assigned.
The icons themselves represent different quality-levels on the level of the source, assuming that contained dataset likely meet those quality levels represented by the source.

\paragraph{Published with Information on Access to the Data}
\begin{compactenum}[(1)]
    \setcounter{enumi}{-1}
    \item No metadata available
    \item Metadata available, but no data access-information available in the metadata
    \item Metadata available, data access-information available only in human-readable form
    \item Metadata available, data access-information available only in human readable form, including general license information
    \item Metadata available, data access-information available in human-readable and  machine-readable form*, including license information
\end{compactenum}

\bigskip\noindent
*The information is available in a structured form that allows, for example, that the respective information can be read out via script.

\subsubsection{Dimension Openness}

\paragraph{General Degree of Openness}
\begin{compactenum}[(1)]
    \setcounter{enumi}{-1}
    \item No information on open accessibility/availability of the data at all
    \item Information available: no open accessibility/availability of the data. No justification, no information on possible contact or restrictions
    \item Like (1) + information on possible contact, restrictions or potential use cases on request available
    \item Like (2) + with justification AND/OR date of moratorium
    \item Open accessibility with corresponding license (no login or contact needed or otherwise with justification)
\end{compactenum}

\pagebreak
\paragraph{Primary Data Formats}
\begin{compactenum}[(1)]
    \setcounter{enumi}{-1}
    \item No primary data available in digital form
    \item Primary data generally available
    \item Primary data stored in common proprietary data formats
    \item Primary data stored in open formats 
    \item Primary data makes use of common, domain specific terminologies (e.g., codelists)
\end{compactenum}

\subsubsection{Dimension Curation}

\paragraph{Level of Curation}
\begin{compactenum}[(1)]
    \setcounter{enumi}{-1}
    \item Data is published in raw form without any curation or documentation (e.g. raw long-tail data)
    \item Data is published in raw form without curation but according to standard with basic documentation like readme (e.g. automatic generated sensor data, long-tail data following a basic scheme)
    \item Data is published in cleaned form with some curation (e.g. brief checking, documentation according to standard)
    \item Data is published in cleaned form with enhanced curation and/ or reprocessing (e.g. conversion to new formats, enhancement of documentation)
    \item Data is published after undergoing extensive curation and/or reprocessing according to discipline specific standards in order to enhance to max. quality (like (3) + additional editing of deposited data for accuracy)
\end{compactenum}

\subsubsection{Dimension Metadata}

\paragraph{Metadata to find/retrieve a Resource / Formal Metadata}
\begin{compactenum}[(1)]
    \setcounter{enumi}{-1}
    \item No metadata available
    \item Metadata available for/with the data publication \textbf{that is not structured} according to a commonly accepted scheme (i.e. no scheme applied)
    \item Metadata provided with the data publication \textbf{that is structured} in a \textbf{basic} way according to a commonly accepted scheme (e.g. completed DataCite mandatory-properties/discovery ; Dublin Core, etc.)
    \item Metadata provided with the data publication \textbf{that is structured} in an \textbf{advanced} way, according to a commonly accepted scheme (e.g., completed Datacite mandatory- and recommended-properties for discovery + discovery-supporting basic content metadata according to DataCite scheme)
    \item \textbf{Full} Metadata provided with the data publication (complete DataCite mandatory- and recommended- and optional-properties for discovery + comprehensive discovery-supporting  content metadata according to DataCite scheme)
\end{compactenum}

\pagebreak
\paragraph{Content related Metadata}
\begin{compactenum}[(1)]
    \setcounter{enumi}{-1}
    \item No content related metadata available
    \item Some content related metadata available, following a (generic) scheme (e.g. DataCite)
    \item Complete content related metadata available following a (generic) scheme (e.g. DataCite)
    \item Some content related metadata available following standardized form or domain specific scheme
    \item Complete and curated content related metadata available following a standardized form and domain specific scheme (see 3)
\end{compactenum}

\subsubsection{Dimension External View}

\paragraph{Score from Domain Specific Fair Assessment Tool}
\begin{compactenum}[(1)]
    \setcounter{enumi}{-1}
    \item 0-20\% Score reached
    \item 21-40\% Score reached
    \item 41-60\% Score reached
    \item 61-80\% Score reached
    \item 81-100\% Score reached
\end{compactenum}

\bigskip\noindent
Remark:
By integrating this dimension, it would be possible to take a domain specific evaluation of a data publication in regards to community best practices and domain specific requirements on metadata and ideally the data itself into account.
This is based on the assumption that in addition to more generalistic FAIR assessment tools currently available, further tools with domain-specific evaluation schemes will be developed.
The rating scales of these tools can then presumably always be converted into a percentage scale.
A rather low weighting of this dimension must be considered in order to minimize the stimulus for gamification.

\section{The Quality Indicator for research software}
\label{sect:software}

\subsection{Dimensions of quality for research software assessment}

Quality of research software has various aspects to be considered.
Depending on the type of software, the field of application and the scientific goals to be followed in using research software, different aspects of research software are being considered as important.
Therefore, a single quality criterion for research software assessment would not be constructive.
To allow a multi-dimensional perspective on research software, the quality assessment being presented in this document is spanning six dimensions of software quality being given by the proposed FAIR-ST framework.

Using the principles for quality assessment, the FAIR principles have to be spelled-out explicitly with respect to research software.
This is given by FAIR4RS~\cite{ChueHong2021}.
As mentioned above, the four FAIR principles have been augmented with two further quality dimensions.
Thus, the six dimensions of research software quality being presented in the assessment scheme are given as follows.

\begin{enumerate}[(a)]
    \item \textbf{Findable}:
        Researchers need to be able to find the software using typical search strategies.
        Having found a given software product, they have to be able to identify a given version/release and be provided with enough information to value the software for their specific research.
    \item \textbf{Accessible}:
        Software needs to be accessible in order to unfold its potential.
        Hereby, accessibility has both a legal side as well as an operational side.
    \item \textbf{Interoperable}:
        For software to be interoperable, it must be capable of being included into larger contexts or frameworks.
        Towards this aim, input/output formats need to be considered as well as interfaces to use the software in automated pipelines.
    \item \textbf{Reusable}:
        Using software for exactly the purpose it has been developed for is considered as accessibility within the quality assessment.
        In contrast, reusability considered the aspect of being able/allowed to use the code for one’s own purposes, to adapt and extend the software.
    \item \textbf{Scientific basis}:
        Research software is an integral part of the research process.
        It therefore has to also follow community/organization specified common standards in performing research.
        While the contribution of software to a certain scientific achievement must be evaluated by dedicated experts in the field, some aspects of scientific quality can also be considered in wider generality.
    \item \textbf{Technical basis}:
        Quality of software also reflects general aspects of software engineering.
        At the end, software has to guarantee to actually produce what it has been aimed to do.
        Professional software development aims at producing software following state-of-the-art software engineering concepts.
        Thus, aspects of professional software engineering must also be considered as part of a quality assessment of research software development. 
\end{enumerate}

The six dimensions of quality for research software are still too generic to be evaluated using the maturity model.
Therefore similar to the approach for research data, each dimension is split into a set of attributes.
The same principles as outlined previously for research data are applied. 

\subsection{Quality assessment of the dimensions}
\label{sect:software:indicators}

The following descriptions summarize the proposed five levels of maturity used for evaluating the quality of software publications.
Level 0 is the default level, when no information for the evaluated attribute is provided.

\begin{enumerate}[(1)]
    \setcounter{enumi}{-1}
    \item Non-existent: no information available
    \item Initial: initial information available being obtained in an ad-hoc, unorganized manner
    \item Repeatable: the information is complete, being produced in a repeatable, yet intuitive manner
    \item Defined: a process is established guaranteeing the complete compilation of the required information
    \item Managed: the process being established is managed, i.e. monitoring/measuring is included
    \item Optimized: practices are put in place optimizing the way the process is operated, leading to improved quality over time
\end{enumerate}

For each of the quality dimensions, the assessment is done by rating every corresponding attribute according to its maturity.
The explicit interpretation of each maturity level for a given attribute is given in terms of asking of compliance with statements.
Since the maturity levels are building on each other in a cumulative way, the number of the last achieved statement defines the maturity of the attribute.

The following sections provide the assessment scheme for the six dimensions of the FAIR-ST framework. 

\subsubsection{Dimension Findable}

The following statements address the aspect of being able to find and uniquely identify the software.

\paragraph{Open Publication Repository}
\begin{compactenum}[(1)]
    \setcounter{enumi}{-1}
    \item There is no information available on where to find the software.
    \item The software is contained in an online repository.
    \item Some kind of description is available giving further information on the software in this repository (e.g. readme file).
    \item A structured meta data description (e.g. following DataCite) given for software is in this repository.
    \item The repository is listed in some overarching meta-repository (e.g. Helmholtz Research Software Directory (RSD)).
    \item The meta-repository is performing quality checks  (e.g. re3data) for the used publication repository.
\end{compactenum}

\paragraph{Versioning}
\begin{compactenum}[(1)]
    \setcounter{enumi}{-1}
    \item No software versioning applied.
    \item There is some kind of versioning for the software.
    \item The software uses structured (e.g. semantic) versioning.
    \item A description of the versioning scheme is available.
    \item There is a documentation on release cycles for the software.
    \item The versioning scheme allows for automatic tagging by CI/CD processes.
\end{compactenum}

\paragraph{Persistent Identifier (PID)}
\begin{compactenum}[(1)]
    \setcounter{enumi}{-1}
    \item No PIDs given.
    \item A handle/URL is provided to identify the software.
    \item The handle/URL is provided with a defined metadata scheme.
    \item A persistent identifier is provided.
    \item A PID allowing for automated harvesting of metadata information is provided.
    \item The PID is part of an established community standard.
\end{compactenum}

\paragraph{Rich Metadata}
\begin{compactenum}[(1)]
    \setcounter{enumi}{-1}
    \item No metadata given.
    \item Some metadata information is provided with the software.
    \item The metadata information is following a given metadata scheme complete.
    \item A metadata curation process reflects changes/updates.
    \item All metadata information following the given metadata scheme can be automatically harvested.
    \item An external quality assessment of the metadata exists.
\end{compactenum}

\subsubsection{Dimension Accessible}

The following questions address the aspect of being able to access research software.
Accessing included the possibility to run the software, which might also be in terms of a web service.
However, accessibility does not include the possibility to adjust the code which is rather being captured under the aspect of reusability. 

\paragraph{Access Conditions (organizational)}
\begin{compactenum}[(1)]
    \setcounter{enumi}{-1}
    \item Not specified.
    \item A contact is given which to inquire about the right to use the software.
    \item The software has a license describing rights of use.
    \item The license allows for open use of the software (e.g. OSI licenses).
    \item There is a way to also obtain some kind of support in using the software.
    \item There isa community, providing the opportunity of support and exchange concerning aspects of using the software.
\end{compactenum}

\paragraph{Access Options (process)}
\begin{compactenum}[(1)]
    \setcounter{enumi}{-1}
    \item There is only one specific form of accessing the software or no option at all.
    \item The software (source code or executable) is provided.
    \item The sources or executables being provided include some documentation on how to install/use the software.
    \item Provided test cases allow to determine whether installation/execution worked as being expected.
    \item Provided checks make sure the software works correctly.
    \item A software service is provided, i.e. are reported bugs taken into the development cycle.
\end{compactenum}

\paragraph{Technical Accessibility (run/start)}
\begin{compactenum}[(1)]
    \setcounter{enumi}{-1}
    \item No information given.
    \item “How to install” information is provided.
    \item Installation scripts are provided.
    \item The software allows for (semi-)automated installation, e.g. a Makefile or manual package (like Python modules).
    \item Sources are provided such that a package manager or automated build tools , e.g. automake, can be used.
    \item A complete package  that enables execution (e.g. container, app package) is available.
\end{compactenum}

\subsubsection{Dimension Interoperable}

The following questions address the aspect of being interoperable, i.e., the possibilities of being able to integrate the software into one’s own software framework or execution pipelines.

\paragraph{Input/Output Formats}
\begin{compactenum}[(1)]
    \setcounter{enumi}{-1}
    \item Not specified.
    \item Some description of input and output formats is provided.
    \item The software builds on standard formats for input and output.
    \item Additional options for varying input/output formats are provided.
    \item The software builds on accepted community standards for input/output data.
    \item The software provides in addition further tools for processing input/output data.
\end{compactenum}

\paragraph{Adaptability/Flexibility of Use}
\begin{compactenum}[(1)]
    \setcounter{enumi}{-1}
    \item No information given.
    \item There is a way to use the software with one defined set of input data.
    \item There are parameters to adjust the way the software is working.
    \item There is some way of logging what is done during execution.
    \item Documented API(s) are provided to integrate the software into one’s own framework.
    \item There is documented way to integrate the software into open workflows, e.g. via containers, web-services etc.
\end{compactenum}

\subsubsection{Dimension Reusable}

The following questions address the aspect of being reusable.
In addition of being accessible, i.e. executable, reusability includes the possibility to actually change/adapt the code. 

\paragraph{Reusability Conditions}
\begin{compactenum}[(1)]
    \setcounter{enumi}{-1}
    \item Not clear.
    \item The software uses  a custom license allowing reuse.
    \item The software uses  a FOSS/OSI approved license including that license dependencies are at least being checked manually.
    \item The software uses  an appropriate license for different file types (code, text, images etc.) following e.g. the REUSE specification.
    \item There is a process available for automatically checking e.g. the REUSE specification.
    \item There is a process available such that all license dependencies are automatically controlled.
\end{compactenum}

\subsubsection{Dimension Scientific basis}

The following questions address the aspect of the software being scientifically well grounded.
While domain specific scientific requirements have to be assessed as part of a scientific peer-review process, certain generic aspect of good scientific practice can be assessed for all research software.

\paragraph{Community Standards}
\begin{compactenum}[(1)]
    \setcounter{enumi}{-1}
    \item No information given.
    \item The connection to known (scientific) standards is drawn.
    \item The software follows standards of the relevant scientific community.
    \item The software complies with relevant scientific standards of the field.
    \item There is an indication on how further evolution of community standards will be addressed.
    \item A closed feedback-loop is established, making sure that further evolutions of community standards are being adopted.
\end{compactenum}

\paragraph{Team Expertise}
\begin{compactenum}[(1)]
    \setcounter{enumi}{-1}
    \item No information given.
    \item Clear expertise from a single, relevant domain is part of the software development team.
    \item The software development team has access to expertise in several relevant domains.
    \item The software development team has access to expertise in all relevant domains.
    \item A fixed, established, interdisciplinary team works on the software.
    \item An established and coordinated community of software developers works on the software.
\end{compactenum}

\paragraph{Scientific Embedding}
\begin{compactenum}[(1)]
    \setcounter{enumi}{-1}
    \item No information given.
    \item At least one scientific use case is documented.
    \item A broader scientific context is documented including several examples.
    \item The software development is at least loosely connected to some scientific initiative.
    \item The software development is part of a larger scientific initiative.
    \item The software development ispart of a larger scientific initiative with dedicated processes for software development.
\end{compactenum}

\subsubsection{Dimension Technical basis}

The following questions address aspects of professional software development leading to sustainable, high quality research software.

\paragraph{Project Management}
\begin{compactenum}[(1)]
    \setcounter{enumi}{-1}
    \item No information on project management and code history being provided.
    \item Some kind of version control is used.
    \item A version control system is used.
    \item A version control system being part of a code project management platform (e.g. GitHub, GitLab) and an associated ticket system is in place.
    \item A transparent process for ticket resolving, code review by other developer, and merge requests is established.
    \item A release process with guaranteed changelog generation, testing, and product provisioning is established.
\end{compactenum}

\pagebreak
\paragraph{Repository Structure}
\begin{compactenum}[(1)]
    \setcounter{enumi}{-1}
    \item No information given.
    \item All files are provided in some structured/unstructured way inside the repository.
    \item The repository is structured albeit maybe in a manner such that every contributor is free to follow own way of organizing files.
    \item A contribution mechanism is documented, e.g. CONTRIBUTORS.md file, as well as a defined structure for the repository and a documented onboarding process.
    \item A common template for the repository structure is available, as well as some kind of identification of deviation.
    \item A repository structure is enforced following community standards.
\end{compactenum}

\paragraph{Code Structure}
\begin{compactenum}[(1)]
    \setcounter{enumi}{-1}
    \item No information given.
    \item Every developer is free to use his/her own style of coding.
    \item There are general recommendations for coding, albeit every developer being able to follow his/her own style.
    \item There is some harmonization of code style being enforced following common standards including meaningful naming of functions/variables etc.
    \item The code style is checked when accepting changes into the repository.
    \item The code style is enforced via a review process (e.g. failed pipelines or auto-formatting).
\end{compactenum}

\paragraph{Reproducibility (Code)}
\begin{compactenum}[(1)]
    \setcounter{enumi}{-1}
    \item No tests, or duplicated code.
    \item The code follows a modular structure allowing for component reusability.
    \item Clear system requirements are documented with min/max versions, albeit version pinning, modularity etc. being enforced manually.
    \item A package manager is used for dependency pinning and testing enforced.
    \item Test coverage is measured, albeit tests may be written on a voluntary basis.
    \item Automated testing for different system environments, requirements for minimal test coverage, and provisioning of containerized packages is done.
\end{compactenum}

\paragraph{Code change process}
\begin{compactenum}[(1)]
    \setcounter{enumi}{-1}
    \item No information 
    \item Internal 4-eye principle for accepting changes
    \item Code changes via transparent processes, e.g. merge/pull request
    \item Approval of code changes via transparent processes and with a 4-eye principle
    \item Integration of code changes into main development branch/releases only allowed for specifically named/trained persons.
    \item Software releases involve an external review (by someone outside of the core developer team)
\end{compactenum}

\pagebreak
\paragraph{Security}
\begin{compactenum}[(1)]
    \setcounter{enumi}{-1}
    \item No security concepts given.
    \item There are at least sporadic updates and dependency checks.
    \item There is a systematic assessment of dependencies and documentation of the software stack.
    \item Deployment is provided within a CI/CD framework for different environments including tools for check for security leaks.
    \item There is some process for monitoring dependency updates including reporting.
    \item There are regular and automated security monitoring and an automated update process in place allowing merges only of security checks have been passed.
\end{compactenum}

\section{Conclusion, Next Steps}

The procedure described so far represents the result of the conceptualization phase of defining a quality indicator for data and software publications as initiated within the Helmholtz Association.
The following main principles were guiding the developing process:

\begin{itemize}
    \item 
        The indicator must match the quality criteria that are essential for the intended objectives, technical issues must not lead to deviation from that target.
        The objectives defined in the Helmholtz Association are: 
        a) to create awareness and appreciation to the diversity of research output, 
        b) to align the assessment of research and research practice to the conditions of digitalization and openness of science, and 
        c) as a stimulus and incentive for scientists to improve their data and software publications related to specific quality criteria, and thus, to improve the quality and trustworthiness of science as a whole.
        Therefore, our basic approach for developing the quality indicator was to define the quality criteria that are essential for the intended objectives in a first step, and to then condense the list of criteria in a second step with respect to technical feasibility of automated collection. 
    \item 
        The indicator should map a multi-dimensional suite of quality criteria onto one value, the quality indicator.
        The approach of solely using simple and one-dimensional metrics to assess research quality and performance is broadly criticized.
        Taking this into consideration the quality indicator for research data and research software publications should provide one easily comparable value that still does more then only counting - by representing a diverse set of quality features in itself. 
    \item 
        The quality indicator has to match all types of research data publications and research software publications generated in the diverse scientific disciplines within the Helmholtz Association, generic approaches are required to provide answers to these questions.
    \item 
        The indicator should rely on generic well-established concepts for quality criteria and quality assessment to make use of already existing preliminary work.
\end{itemize}

The result of the conceptual development process is a generic multi-dimensional quality indicator based on a defined set of goal-oriented quality dimensions and attributes as well as well-established, maturity-based concepts for quality measurement.

Feasibility of the concept is currently evaluated in pilot implementations and further refinements.
Several aspects have to be clarified in the following piloting phase of the implementation of the proposed procedure.

\begin{enumerate}[(i)]
    \item 
        While the attributes for each quality dimension have been discussed and iterated in several rounds of community engagement, their actual application to assess the quality of given data or software will show whether the formulation of the attributes is clear enough to be applied in a wider community, where documentation of the assessment needs to be improved, or where even manual assessment is not easily possible at the moment.
    \item
        In particular, the applicability of the quality indicators in various research communities needs to be proven by running a couple of exemplifying assessments.
    \item 
        The quality assessment should only lead to a minimal increase in the resources required for the assessment itself.
        Therefore, the assessment of all the quality dimensions and attributes should be automatized as much as possible.
        Clearly, working with repositories for research data and research software is a key requirement for automation on a higher level.
        Therefore, introducing quality indicators for data and software publications hopefully further fosters the level of professionalization in dealing with these outcomes of scientific work.
    \item
        The final assessment and visualization should be done in a digital manner. Suitable tools are currently assessed or being developed.
\end{enumerate}

It is obvious that the described approach to derive multi-dimensional quality indicators can be used in an analog way to assess the research endeavor as a whole.
In the end, the adoption of a precise definition of these indicators, the fine-tuned weighting scheme of quality dimensions and attributes against each other, as well as the definition of a minimal level of maturity to be expected for research data and research software, will make the process of research assessment transparent and open.
Moving towards multi-dimensional assessments rather than single numbers will better reflect the complexity and diversity of modern scientific work and, at the end, lead to a fairer assessment of the contributions of the various parts of the scientific system towards the common goal of contributing to increase collective human knowledge. 

With this paper, we share the current state of the process within the Helmholtz Association.
We invite all interested parties to further expand, refine, or change the dimensions and attributes being proposed here.
It is our belief that any list of criteria has to be a living document to account for the dynamic nature of research.
Only through constant evolution, the proposed framework is able to keep up with new thoughts, ideas, and findings on how to improve the scientific process.

%% file: tex/2_acknowledgment.tex
This paper reflects the work of the Task Group Helmholtz Quality Indicators for Data and Software Publications of the Helmholtz Association from the past two years.
While not all group members were able to participate in this paper, the authors acknowledge and highly appreciate the constructive discussions of the whole group, namely 
Gisbert Breitbach~(Hereon),
Torsten Bronger~(FZJ),
Stefanie Castell~(HZI),
Claas Faber~(GEOMAR),
Tamara Husch~(HZB),
Ulrike Kleeberg~(Hereon),
Matthias Kretz~(GSI),
Marco Nolden~(DKFZ),
Maria Nüchter~(KIT),
Florian Ott~(GFZ),
David Pfaff~(CISPA),
Will Rayner~(HMGU),
Marco Schaefer-Herte~(DZNE),
Dagmar Sitek~(DKFZ),
Christian Tacke~(GSI),
Anne Talk~(HZI), 
Regine Tobias~(KIT), and
Andreas Walker~(AWI).

An initial version of the paper was also shared with research software and data experts within the Helmholtz Association.
The authors acknowledge and also incorporated comments to these early versions from 
Stephan Druskat~(DLR), 
Tobias Huste~(HZDR), 
Sven Kiele~(HZDR),
Uwe Konrad (HZDR), 
Tobias Schlauch~(DLR), 
Bernard Silenou~(HZI), and 
Thomas White~(DESY).

%% file: tex/3_credit.tex
\section*{CRediT author statement}

\begin{compactdesc}
    \item[Wolfgang zu Castell] Conceptualization, Methodology, Writing – Original Draft 
    \item[Doris Dransch] Conceptualization, Investigation, Writing – Review and Editing, Supervision, Project administration
    \item[Guido Juckeland] Conceptualization, Writing – Review and Editing, Visualization, \newline Supervision
    \item[Marcel Meistring] Conceptualization, Writing – Review and Editing, Supervision, \newline Project administration
    \item[Bernadette Fritzsch] Conceptualization, Writing – Review and Editing
    \item[Ronny Gey] Conceptualization, Writing – Review and Editing
    \item[Britta Höpfner] Conceptualization, Writing – Review and Editing
    \item[Martin Köhler] Conceptualization, Writing – Review and Editing
    \item[Christian Meeßen] Conceptualization, Investigation, Writing – Original Draft
    \item[Hela Mehrtens] Conceptualization, Writing – Review and Editing
    \item[Felix Mühlbauer] Conceptualization, Investigation, Writing – Original Draft
    \item[Sirko Schindler] Conceptualization, Writing – Review and Editing
    \item[Thomas Schnicke] Conceptualization, Writing – Review and Editing
    \item[Roland Bertelmann] Supervision, Project administration
\end{compactdesc}